# Design criteria of flexible capacitive pressure sensors using DIY-techniques and household materials

Thesis submitted in partial fulfillment
of the requirements for the degree of

*(Master of Science in Electronics and Communication Engineering by Research)*

by

Rishabh Bhooshan Mishra
2018702005
rishabh.bhooshan@research.iiit.ac.in

InternationalInstituteofInformationTechnology, Hyderabad
(Deemed to be University)
Hyderabad-500032,INDIA

May – 2021

InternationalInstituteofInformationTechnology
Hyderabad,India

# Certificate

It is certified that the work contained in this thesis, titled “Design criteria of flexible capacitive pressure sensors using DIY-techniques and household materials” by Rishabh Bhooshan Mishra, has been carried out under my supervision and is not submitted elsewhere for a degree.

20th May 2020

Adviser: Dr. Aftab M. Hussain

ToMy Parents

## Acknowledgments

I thank my dissertationsupervisor Dr. Aftab M. Hussain for considering me in his group and helping me in enhancing my research qualities. Apart from being an accomplished trained supervisor from IIT – Roorkee, KAUST, and Harvard University, he is an amazing human being and friend. He always gave me full freedom to explore my strength, helped me overcome my weakness, and provided me moral support to pursue my further studies.

A special thanks to Prof. Muhammad M. Hussain [presently affiliated with Electrical and Computer Sciences, University of California – Berkeley and Electrical and Computer Engineering, King Abdullah University of Science and Technology (KAUST)] who considered me as visiting student in MMH Labs, Electrical and Computer Engineering at KAUST during Fall-2019, Summer-2020, Fall-202,0 and Spring-2021. Prof. Hussain provided me so many opportunities to work on various projects of his group. Apart from being an amazing supervisor, he is an amazing and commendable manager which I experienced during Fall-2019 when I was working under him at KAUST. He is still training me remotely in this global pandemic COVID-19 and provides support, guidance, care, and inspiration.

I would like to thank my MS dissertation committee members for considering my thesis to review and giving some time from their busy schedule. I would like to thank Dr. Nazek El-Atab (currently Research Scientist and upcoming Assistant Professor at Electrical and Computer Engineering at KAUST], Dr. Nadeem Qaiser (currently Research Scientist at KAUST), Dr. Sherjeel M. Khan [currently Research Scientist, Silicon Austria Labs] and Dr. Sohail F. Shaikh [Postdoctoral Researcher, IMEC]for being my mentors since my first internship (from Fall-2019) at KAUST who helped me a lotwith the fabrication and characterizations of sensors. I would like to thank my friends Wedyan Babatain [currently PhD student at KAUST], Uttam Kumar Das [currently PhD student at KAUST],Anis Fatema [currently PhD student at IIIT-Hyderabad], and Sumana Bhattacharji [currently MS student at IIIT-Hyderabad and upcoming PhD student at KAUST], Reem Alshanbari [currently MS student at KAUST and upcoming PhD student at Columbia University], and Sandeep Nagar [currently MS student at IIIT-Hyderabad]for their constant help, support, and encouragement with technical and non-technical discussion. Finally, I would like to thank my parents from bottom of my heart for their limitless patience and support. Special thanks to my siblings for always being with me in all of my decisions.

# Abstract

The flexible capacitive pressure sensors are one of the most essential and famous devices with vast applications in automobile, aerospace, marine, healthcare, wearables, consumer, and portable electronics. The fabrication of pressure sensors in a cleanroom is expensive and time-consuming; however, the sensitivity, linearity, and other performance factors of those pressure sensors are exceptional. Moreover, sometimes we require sensors that are not expensive and can be fabricated rapidly where the other performance factors don't need to be highly remarkable. In this modern era, household materials and DIY (Do-it-yourself) techniques are quite helpful, highly utilized. They are recommended to fabricate low-cost sensors and healthcare devices for personalized medicine and low-cost consumer electronics. Different flexible capacitive pressure sensors are presented and experimentally characterized for acoustic and air-pressure monitoring in this thesis. The design criteria of a cantilever-based capacitive pressure sensor are discussed. The three different designs are analysed with aspect ratios of 1.5, 1.0, and 0.67. The sensor with an aspect ratio of 0.67 shows maximum sensitivity (mechanical and electrical), better response time, and the 1st and 2nd mode of resonant frequencies is comparatively less than the other two. The cantilever designs are susceptible to slight pressure; therefore, the diaphragm-based normal mode capacitive pressure sensor is introduced in the second chapter, which defines the design criteria of diaphragm shapes. The five different diaphragms analysed are circular, elliptical, pentagon, square, and rectangular shapes. The circular capacitive pressure sensor shows maximum sensitivity, however, maximum non-linear response. The rectangular capacitive pressure sensor has a maximum linear response; however, the sensitivity is the lowest. However, to design the pressure sensor for bit higher pressure monitoring and increase the linear output regime, we analyse the single and double touch mode capacitive pressure sensors. The single-touch mode capacitive pressure sensor shows linear response for 8 – 40 kPa however double touch mode capacitive pressure sensor shows the linear response for 14.24 – 54.9 kPa. The presented sensor offers high sensitivity and an accurate response; therefore, we can utilize these sensors for various applications like tire-pressure monitoring, hear-rate, asthma monitoring, etc. The capacitive pressure sensor is chosen because of its advantages like low-temperature drift, high sensitivity, and ease of design.

# Table of contents

# List of figures

# List of tables

# Chapter - 1

# Introduction

## 1.1 Intro to MEMS and capacitive pressure micro-sensors

The MEMS (Micro-Electro-Mechanical System) is one of the most popular areas in electrical and electronics engineering, which broadly combines various physical, chemical, and biological processes or phenomena on a chip. The MEMS research area became more popular after process advancement in complementary-metal-oxide-semiconductor (CMOS) and Integrated Circuit (IC), which combines multiple microfabrication techniques and micromachining. All different MEMS devices (except microfluidic sensors and devices) consist of mechanical components mechanical which show deformation or movement after physical process parameters applications [1]–[10].

The MEMS-based pressure sensors are an essential device in the micro-sensor arena, covering a maximum percentage of the market for different industries for pressure parameter monitoring. Multiple design approaches and principles are being used to monitor different types of pressures, i.e., absolute, gauge and differential.

- *Differential pressure* is the difference between two pressure which is technically w.r.t. another pressure
- *Absolute pressure* is being measured w.r.t. vacuum
- *Gauge pressure* is being measured w.r.t. atmospheric pressure

These three different types of pressure have been monitored previously in pervasive and advanced ways using piezoresistive [2], capacitive [1], resonant [11], and optical [12] techniques.

The piezoresistive sensing is most utilized for sensing technique for pressure monitoring among all four types mentioned above. The MEMS piezoresistive pressure sensors, the piezoresistive element, are placed at the micro-mechanical component's highly stressed region, which is sensitive to pressure application. The deflection in the mechanically sensitive part causes stress, and the piezoresistive details, which are placed in this region, change resistivity that defines applied pressure. Measuring the change in resistivity is performed using the Wheatstone Bridge method, which monitors the change in resistance when the bride unbalances due to pressure application [2], [13], [14]. The piezoresistive pressure sensor is highly applicable to monitor change in pressure with high sensitivity, provides linear response for extended dynamic range, high reliability, and small size [2], [13], [14]. However, mass fabrication and temperature sensitivity are the significant problems with the piezoresistive pressure sensors, which capacitive pressure microsensors have covered. On the other hand, the capacitive pressure sensors offer high sensitivity, precise measurement, and less temperature drift [2], [13], [14].

## 1.2 Why MEMS to flexible

MEMS-based devices are popular due to their outstanding performances and high reliability;

however, they are rigid and can't be applicable for non-curvilinear surfaces. The freeform (physically flexible, stretchable, and reconfigurable) CMOS electronics are becoming in trend to make devices flexible, stretchable and expandable, which is not only limited to control, computation, communication and display, healthcare, and Internet of Everything (IoE) [15]–[22] even it is utilized for killer applications, vehicular technology, marine ecology, harsh environment applications and solar cells as well [15], [17], [20]–[26]. The freeform CMOS enabled electronics to become popular due to the technological advancement and requirement of placing the electronics on soft surfaces (like human skin or skin of other species and plants) for various environmental monitoring. This technology revolutionized healthcare with wearables and implantable electronics (glaucoma monitoring and brain-machine interface). To overcome with the multiple challenges of contemporary electronics, the freeform CMOS electronics became popular however it requires some cleanroom processing which increases the cost of the device and requires time for mass fabrication; therefore, some other fabrication technique is being adapted like rapid/additive manufacturing, DIY and garage fabrication techniques.

## 1.3 DIY-technique and paper as active material

Design, fabrication, and performance enhancement of low-cost MEMS-based pressure sensors are in trend in the last few years. The development of low-cost MEMS-based devices is significant in the modern era to reduce the cost of devices without performance degradation. However, sometimes the devices' performance is as essential as the cost-effectiveness; therefore, device fabrication from household materials using DIY-technique became very popular, and paper electronics evolved. The paper-like materials and other household materials became very useful to design the sensors [27]–[32], actuators [33], [34], microfluidic devices [35], and transistors [36], [37].

The DIY technique is becoming popular because it is beneficial for rapid manufacturing and helps to design low-cost flexible and printed electronic systems. These techniques suggest that everyone in this world could be able to make their electronics device [38]. D. A. Mellis from MIT Media Lab presented DIY based radio, speaker, cellphones, and mouses from low-cost materials, which gave a new introduction to digital fabrication, embedded and passive computation [38]. J. M. Nassar et al. presented pressure, temperature, and humidity sensors from household materials like Al-foil, double-sided tape, a microfiber wipe, conductive ink, and sponge and then designed the electronic skin [29] and paper watch out of these materials [39]. The pressure sensor is designed after using the capacitive pressure sensing principle. Al-foil acts as parallel plate electrodes; however, air, a microfiber wipe, sponge, and double-sided tape are utilized as dielectric materials. In the capacitive pressure sensing principle, the mechanically sensitive element (i.e., Al-foil in this case) deflects, which causes the change in the separation gap. That shift in the separation gap causes a change in capacitance. Moreover, that change in capacitance provides information about applied pressure. After analyzing these dielectric materials, the pressure sensor with air dielectric shows the highest sensitivity however saturates faster than others [29]. The pressure sensor of Al-foil with microfiber wipe and double-sided tape as dielectric materials are utilized for pulse rate monitoring in another work of J. M. Nassar et al. [39]. Moreover, S. M. Khan et al. from MMH Labs at KAUST presented the pill counter for personalized medicine/healthcare [40]. The pill counter consists of anisotropic conductive tape with the silver particle in between and sandwiched between two Cu electrodes and characterized for 0 – 40 kPa pressure.

## 1.4 Motivation

Since the microfabrication technique requires cleanroom facilities because the devices are very time-consuming and expensive, to analyze the multiple types and modes of the capacitive pressure sensor, we have utilized household materials and DIY-technique, a.k.a. garage fabrication process.

Household materials such as paper like materials, printable ink, Al-foil, and tapes (single-sided, double-sided and posted) are utilized to design and fabricate multiple electronic devices. The advantages of these devices are:

- cost-effective
- simple to fabricate
- ease in mass fabrication
- no need for specific raw materials, fabrication tools, and techniques

However, herein in this thesis, I present multiple modes, structures, and design analysis of capacitive pressure sensors after noticing these various advantages. The cantilevers/beams of different aspect ratios and diaphragms shapes are utilized for analysis. The different modes of capacitive (normal, touch, and double touch) are also characterized. The Al-coated Kapton sheet, double-sided tape, and Kapton sheet are utilized for sensor fabrication and then characterized. The finite element simulators [COMSOL Multiphysics and CoventorWare®]being used for sensitivity analysis, and mathematical explanations are also presented. In chapter 2, I have analyzed cantilever-based normal mode capacitive pressure sensors of three different aspect ratios; however, the overlapping area between all three designs is kept the same. In chapter 3, five different diaphragm shapes, i.e., circular, square, rectangular, pentagon, and elliptical, are analyzed for normal mode capacitive pressure sensors. In chapter 4, single-touch and double-touch mode, capacitive pressure sensors are presented and characterized.

The contents of all these chapters are published in the following research articles:

- R. B. Mishra, N. El-Atab, A. M. Hussain, M. M. Hussain, "Recent Progress on Flexible Capacitive

Pressures Sensors: From Design & Materials to Applications," *Adv. Mater. Technol.*, vol. 6, no.4, p. 2001023. [Link] [Cover article]

- R. B. Mishra, S. F. Shaikh, A. M. Hussain, M. M. Hussain, "Metal coated polymer and paper-based cantilever design and analysis for acoustic pressure sensing," *AIP Adv.*,vol. 10, no. 5, p. 055112,2020.[Link]
- S. M. Khan*, R. B. Mishra*, N. Qaiser, A. M. Hussain, M. M. Hussain, "Diaphragm shape effect on the performance of foil-based capacitive pressure sensors," *AIP Adv.*,vol. 10, no. , p. 015009, 2020. (*equal first author) [Link]
- R. B. Mishra, W. Babatain, N. El-Atab, A. M. Hussain, M. M. Hussain, "Polymer/paper-based double touch mode capacitive pressure sensing element for wireless control of the robotic arm,"*15th IEEE International Conference on Nano/Micro Engineered Molecular Systems (IEEE-NEMS)*, pp. 95-99, San Diego,CA, USA, 2020. [Link]
- R. B. Mishra, S. M. Khan, S. F. Shaikh, A. M. Hussain, M. M. Hussain, "Low-cost foil/paperbased touch mode pressure sensing element as artificial skin module for prosthetic hand," *3rd International Conference on Soft Robotics (RoboSoft)*, pp. 194-200, Yale University,CT, USA, 2020. [Link]
- R. B. Mishra, S. R. Nagireddy, S. Bhattacharjee, A. M. Hussain, "Theoretical Modelling and Numerical Simulation of Elliptical Capacitive Pressure Microsensor," *2nd IEEEConference on Modeling of Systems Circuits and Devices (MOS-AK India)*, pp. 17-22,Hyderabad, India, 2019. [Link]

# Chapter – 2

# Cantilevercapacitivepressure sensors

## 2.1 Introduction

Cantilevers are one of basic element in MEMS technology for pressure sensors [41], microphones [42], flow sensors [43], gyroscope [44], accelerometers [45], energy harvester [46], resonators [47], grippers [48] and ultrasonic transducers [49]. These devices/sensors utilize different sensing techniques i.e. piezoresistive [50], piezoelectric [46], capacitive [41], electrothermal [51], chemical and biological [52]. Among all these techniques the capacitive one using cantilevers is most utilized among all for automotive [53], aerospace [54], robotics [55], industries (chemical and biological both) [52], consumer and portable electronics [42], [43], [56].

Proceeding with this approach of research, techniques of Do-it-yourself (DIY) like folding, printing, and cutting [28], [29], [33], [50], [57], [58] for designing and analysis of cantilever capacitive pressure sensor is presented for various applications using paper and polymer composites, e.g. Al-foil, Cu-foil, Kapton-tape, metal-coated polymers sheets, scotch tape, and glass. Paper-based cantilever pressure sensors, using piezoresistive sensing, are fabricated from paper, carbon, and silver ink. A square-shaped diaphragm, which is clamped at all edges using four cantilevers, presents a weighing machine [50]. Hygroxpensive electrothermal paper actuators (HEPAs) of a different type; straight, recurved, and released, which is fabricated using paper, conducting polymer (PEDOT: PSS), and adhesive tape, operate due to change in resistance or dimensional parameter when the cellulose paper absorbs humidity/moisture [59]. Laser-induced graphene (LIG) is printed on Kapton polymer sheets which linearized for pressure measurement of extensive dynamic range (20 MPa), which have a sensitivity of 1.23 Pa and resolution of 10 Pa with extremely excellent long-term stability of a minimum 1of 5,000 measurement cycles [60].

In this chapter, Al coated Kapton (PI) sheet, scotch tape, and glass sheet piece are used for designing, fabricating, and analysis is presented. The pressure sensor's sensitive element and backplates manufactured using the Al coated Kapton foil, scotch tape is used to clamp one edge of the cantilever, and the backplate is fixed at a glass sheet piece. The resonant frequency, response time, stability of the system, mechanical and capacitive sensitivity after acoustic/sound pressureapplication.

## 2.2 Governing Equations

Kirchhoff's plate theory for the thin or thick mechanically sensitive element is a unique and remarkable method to the mathematical analysis of sensors which is very much utilized in designing micro and nano-sensors. The deflection in mechanically sensitive components has a significant interest and influence on sensors/devices' static and dynamic performance/behavior of sensors/devices. The large deflection theory is being utilized for mathematical analysis of deflection, mechanical and cantilever sensitivity of cantilever capacitive pressure sensor for acoustic pressure measurement which will generate 1 Pa pressure, in this present chapter. The cantilever after pressure application will follow a large deflection theory, due to which the dynamic behavior of the sensor will be non-linear.

The non-linear partial differential equation using Euler-Bernoulli's theory for deflation in

cantilever is given by:

$$\frac{\partial^4 W(x,t)}{\partial x^4} + m\frac{\partial^2 W(x,t)}{\partial t^2} + q(x,t) = 0 \colon \forall\, x \in (0,L),\ t = 0 \qquad (2.1)$$

where, D, L, m, and qare flexural rigidity of pressure-sensitive cantilever, length of the cantilever, mass per unit length of the cantilever, and distributed load on the cantilever, respectively.

For this initial boundary value problem, the initial conditions are:

$$(x,t) = 0\ \& \frac{\partial W(x,0)}{\partial t} = 0;\ \forall\, x \in [0,L] \qquad (2.2)$$

$$W(x,t) = 0\ \& \frac{\partial W(x,0)}{\partial t} = 0;\ \forall\, x \in [0,L] \qquad (2.3)$$

$$W(0,t) = 0,\ \ \frac{\partial W(0,t)}{\partial t} = 0,\ \ \frac{\partial^2 W(L,t)}{\partial t^2} = 0\ \ \&\ \ -D\frac{\partial^3 W(L,t)}{\partial t^3} = P(t);\ \forall\, t \geq 0 \quad (2.4)$$

Since the cantilever is non-stretchable:

$$\int_0^{\tilde{x}(t)} \sqrt{1 + \left(\frac{\partial W(x,t)}{\partial t}\right)^2}\, dx = L \text{ and } q(x,t) = 0;\ \forall\, t \geq 0 \qquad (2.5)$$

The deflection in cantilever due to pressure wave, $P(t) = P_n \sin(\omega t)$, can be given by:

$$W(x,t) = \frac{P(t)}{2D\beta^3[1 + (\cos\beta L)(\cosh\beta L)]}[\sin\beta(x-L) + \sinh\beta(L-x)$$

$$-\cos\beta x \sinh\beta L + \sin\beta L \cosh\beta x + \sin\beta x \cosh\beta L - \cos\beta L \sinh\beta x] \qquad (2.6)$$

where, $\beta = \sqrt[4]{m\omega^2/D}$ and $\omega$is frequency of vibration.

And the bending moment of the cantilever at the clamped edge can be given by:

$$M(t) = D\left|\frac{\partial^2 w(x,t)}{\partial^2 x}\right|_{x=0} \qquad (2.7)$$

The deflection due to pressure application changes the capacitance is given by:

$$C = \iint_S \frac{\varepsilon \, dA}{d - W} \tag{2.8}$$

Finite Finite element analysis of deflection in cantilever due to pressure application required before fabrication helps to choose proper sensor dimensions, i.e. separation gap between electrodes, the overlapping area between parallel plates, and clamping of edges, which reduces material wastage, leading to minimise sensor's cost and time consumption.

Three different cantilever designs [D1, D2, and D3] of aspect ratios 27.27, 1.0, and 0.0366 have been chosen **[Figure 1-(a)]** for FEM simulations. The CoventorWare® software is used for finite element analysis in deflection in cantilever **[Figure 1-(b)]**, resonant frequency **[Figure 1-(c)]**, mechanical and capacitive sensitivity **[Figure 1-(d)]** with the proper trend in all these parameters as there are increasing in aspect ratios. The thickness of all designs [D1, D2, and D3], overlapping area, and separation gap between both electrodes are 25 µm, 0.15 cm2, and 250 µm, respectively. The D1 and D3 sensor designs are alike, with only one difference in the clamped edge.

The resonant frequency decreases as the aspect ratio decreases, i.e. deflection in cantilever increases which is minimum for design D3 **[Figure 1-(c)]**. The mechanical and capacitive sensitivity increases as the aspect ratio decreases, which is maximum for design D3**[Figure 1-(d)]**.

## 2.4 Fabrication

The cantilever, which is fabricated using our approach, requires DIY-technique or garage fabrication process and low-cost materials like Al coted Kapton foil (Liren's LR-PI 100AM of 25 ¬µm polyimide coated with 200 nm aluminium), double-sided scotch tape, and glass pieces (7.5 cm×5 cm×0.5 cm). The steps to fabricate sensors is as follows:

- **Step-1:** The Al coted Kapton sheet is adhered to on top glass substrate, which will act as carrier substrate to fix the bottom electrode and provides mechanical strength to conductive cantilever structure **[Figure 2-(a, b)]**.
- **Step-2:** To cut the cantilever designs, the Carbon Dioxide (CO2) laser tool scribes the foil, making different patterns of cantilevers from a single carrier substrate. After removing the accessive film, the various designs of cantilevers are obtained **[Figure 2-(c, d)]**. The power, speed, vertical separation between laser and substrate, and the frequency is optimized for scribing, which is 10%, 20%, 2 mm, and 1000 ppm, respectively.
- **Step-3:** Paper adhesive tape is used to clamp the one edge of the cantilever, and other edges are hanging, and the air is the dielectric medium between parallel plate capacitor **[Figure 2-(e, f)]**. The photographic image of the actual sensor is shown in **[Figure 2-(g)]**.

## 2.5 Characterization

### 2.5.1 Experimental set-up

The experimental setup is prepared for acoustic pressure sensing. The fabricated pressure sensor is placed underneath the Bluetooth speaker (JBL Go Portable Speaker) and connected to the Keithley Semiconductor Characterization System (Model – 4200 SCS) for measuring capacitance change/variation after the deflection in the mechanically sensitive cantilever diaphragm. In addition, the Bluetooth speaker is connected to the source (Moto one power mobile), which can play the sound of different frequencies.

## 2.3 FEM-simulations

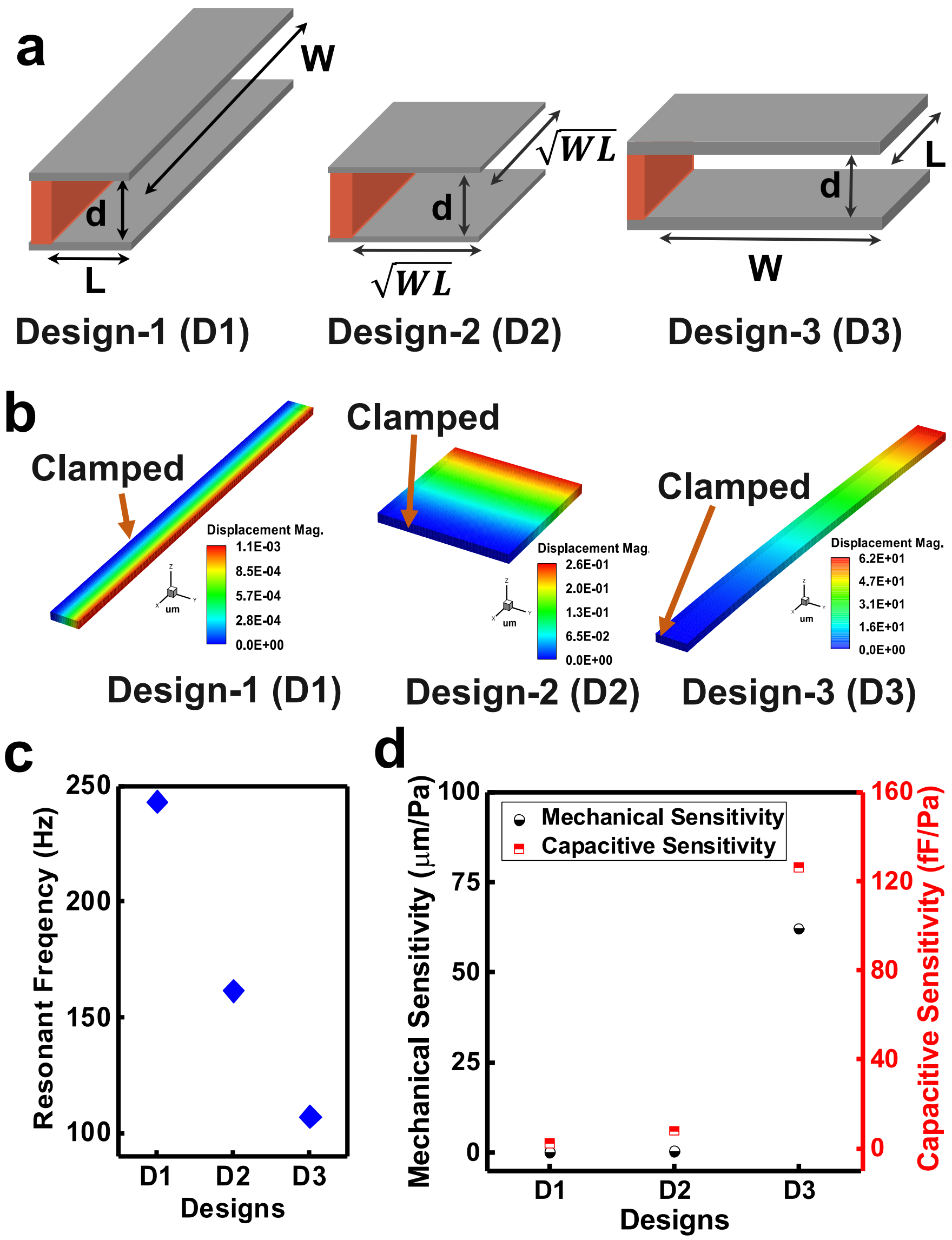

**Figure 1**Geometrical explanation and Finite Element simulations of sensor designs. (a) Schematic of three different designs of cantilever capacitive pressure sensors. The surface area of the diaphragm is the same for all three designs D1, D2, and D3. (b) Deflection in the diaphragm at 1 Pa pressure application. (c) The resonant frequency of all three designs using finite element simulation. (d) Mechanical and Capacitive sensitivity of all three designs using finite element simulation at 1 Pa pressure.

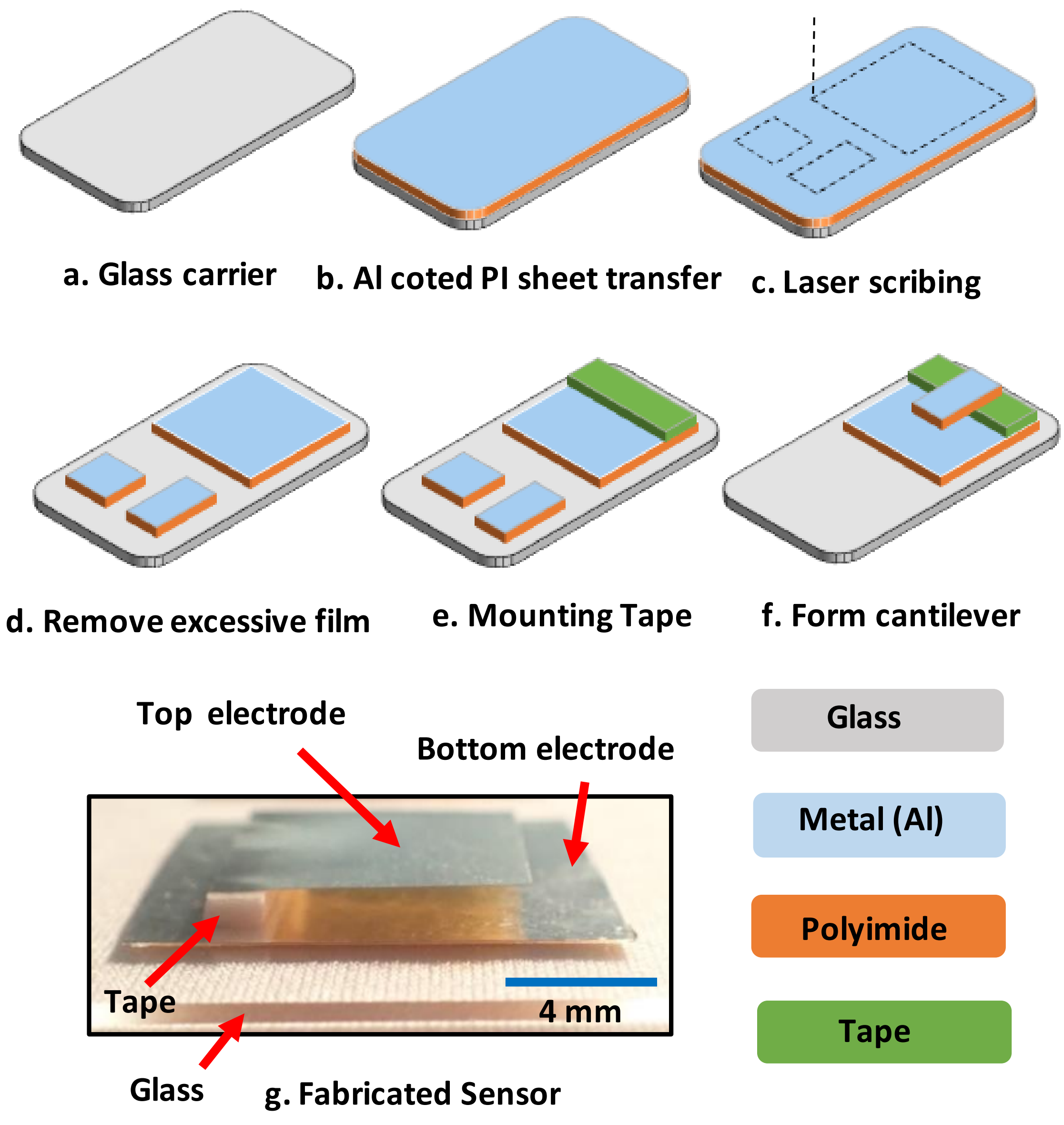

**Figure 2(a-e)** Schematic illustration of steps involved to fabricate cantilevers capacitive pressure sensor in which **(f)** represents the final architecture of sensor. **(g)** Digital photograph of the fabricated capacitive pressure sensor to show the air as a dielectric layer (separation gap) which varies due to sound pressure application.

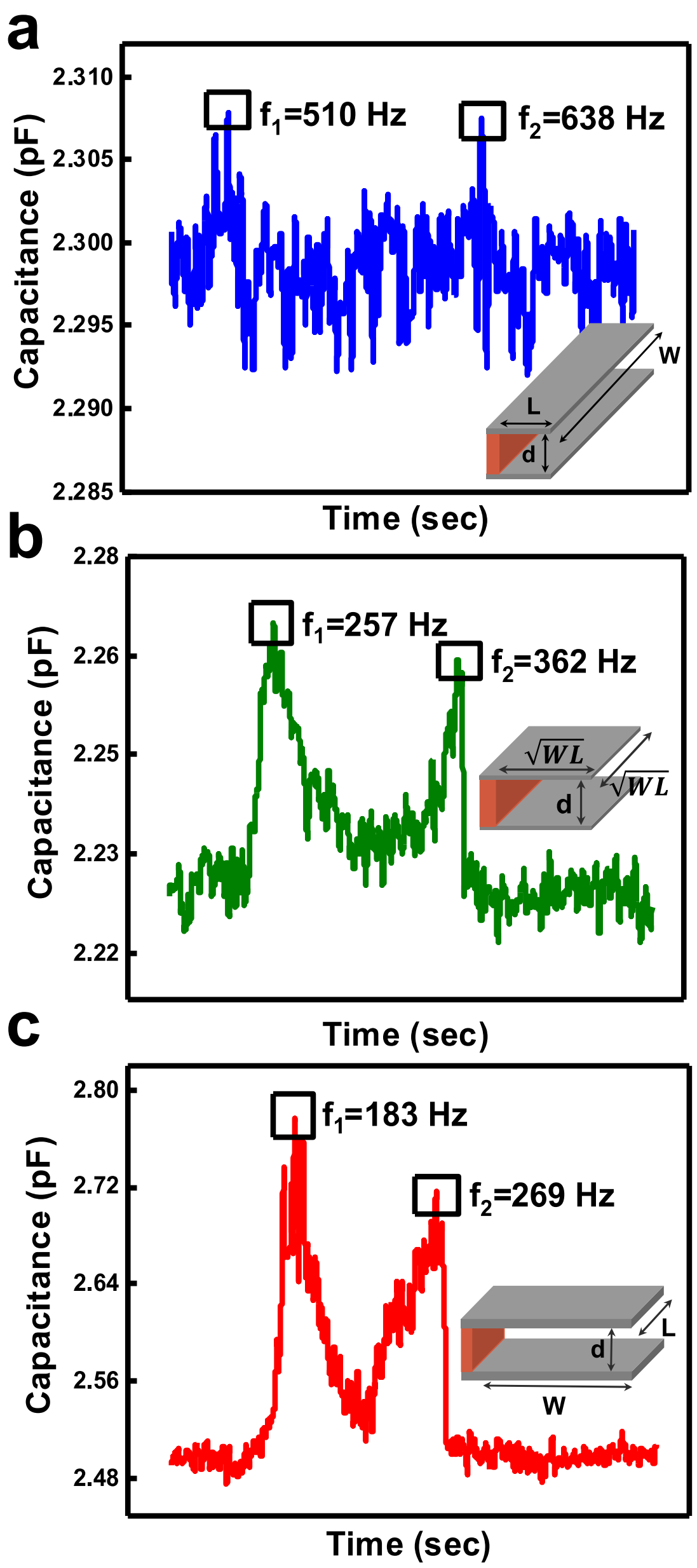

**Figure 3**Experimental results of 1st and 2nd mode of resonant frequency (f1 and f2) after applying the sweep of 20 Hz to 20 kHz acoustic wave for three different cantilever sensor's aspect ratios of (a) 1.5, (b) 1.0, and (c) 0.67. All three cantilevers have the same overlapping area of 1.5 $cm^2$. The first mode resonant frequency is known as the natural frequency which is minimum for the cantilever of the smallest aspect ratio.

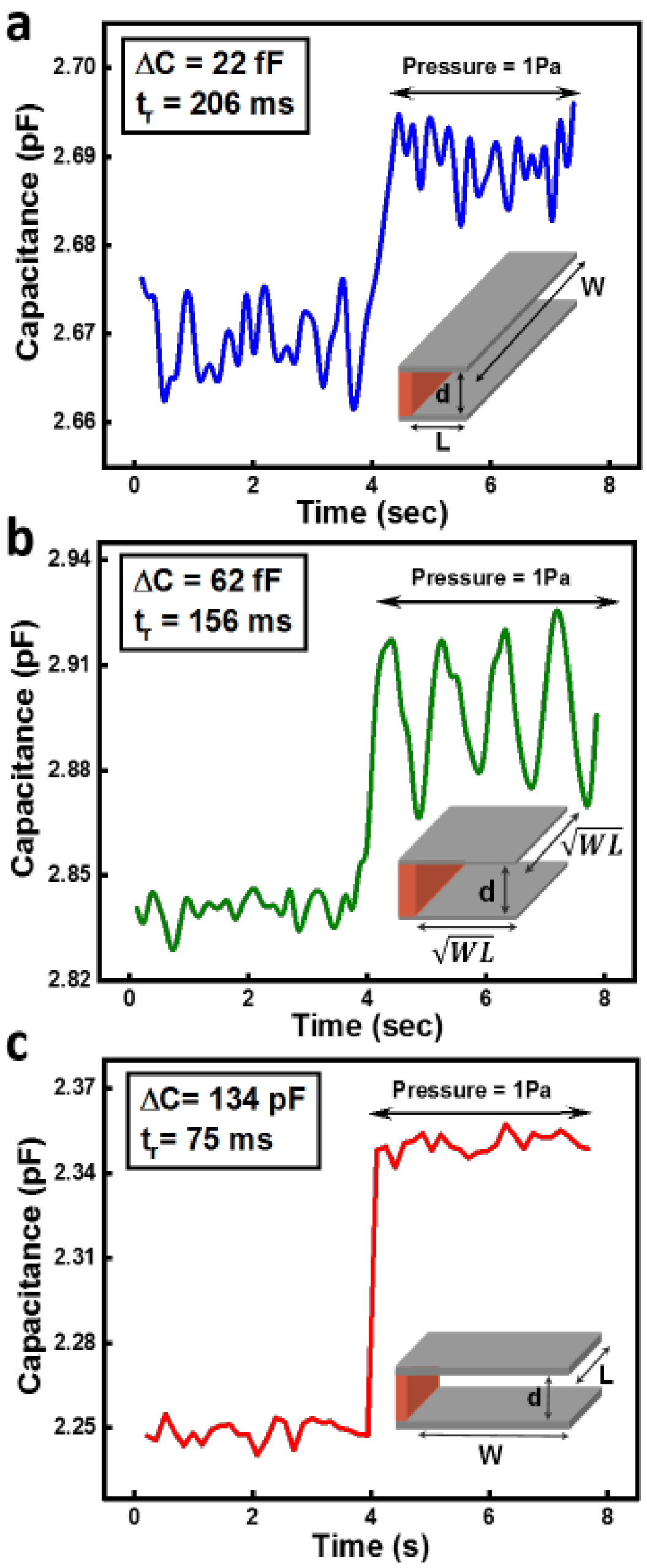

**Figure 4**Experimental results of capacitance change and response time of all three different cantilever pressure sensors which have the aspect ratio of **(a)** 1.5. **(b)** 1.0 and **(c)** 0.67. The sound wave of 300 Hz frequency is applied which is 1 Pa pressure. The cantilever pressure sensor of the smallest aspect ratio has maximum sensitivity and minimum response time among all three designs.

### 2.5.2 Experimental results

A frequency sweep from 20 Hz – 20 kHz is given on all three designs for obtaining resonant frequencies in the designed acoustic pressure measuring setup. The first and second mode of frequencies (f1 and f2) is obtained for all different design of cantilever pressure sensors **[Figure 3-(a-c)]**. The aspect ratio for design D3 is minimum; therefore, the f1 and f2 are less, and the change in capacitance is maximum for the design with the smallest aspect ratio among all designs. It is observed that the experimental result trends replicate the trends in all three designs from finite element analysis. The resonant frequency is the smallest in the design, which has the smallest aspect ratio among all designs and increases as the aspect ratio of the cantilever increases **[Figure 3-(a-c)]**.

We observed that the design which has the smallest aspect ratio has the smallest gap between the occurrence of the first and second mode resonant frequencies ($f2 - f1$) among all designs, due to which this design can be considered as the most stable system among all. The capacitance change in all three designs at a frequency of 300 Hz **[Figure 4-(a-c)]**. 300 Hz frequency is chosen to apply sound pressure because it does not match the resonant frequency of any design. The sound pressure deflects the diaphragm of the design, which has the smallest aspect ratio among all and follows the same trend shown in finite element simulations for all designs. The response time (tr) of the sensor increases as the aspect ratio of the cantilever and minimum rise time for the cantilever sensor of the smallest aspect ratio decrease **[Figure 4-(a-c)]**.

## 2.6 Conclusion

Paper/polymer/foil-based sensing materials have gained significant importance in this emerging electronics area. Herein, we have presented a metal-coated polymer-based cantilever pressure sensor for acoustic pressure sensing using a simple garage fabrication and DIY approach with radially available raw ingredients. The analysis provides an insight on how the geometrical parameters play an essential role for any cantilever pressure sensor and what design shall be preferred based on the frequency spectrum according to applications. After analyzing all three different cantilever designs whose aspect ratios are 0.67, 1, and 1.5, we conclude, the capacitive pressure sensor, which has a maximum aspect ratio, gives a rapid response, possesses maximum sensitivity, and is applicable to respond to low-frequency sound (f2 – f1=183 Hz). However, the sensitivity and response time decrease as the aspect ratio gets lower. Furthermore, the first and second mode of resonant frequencies is undershot for the sensor of maximum aspect ratio, which is advantageous for designing the stable system, which decreases as aspect ratios get lower. Furthermore, more studies according to applications can be performed in the future, such as human health monitoring, flow-sensing, and environmental monitoring.

# Chapter – 3

# Diaphragm based capacitive pressure sensors

## 3.1 Introduction

Apart from cantilevers, different diaphragm shapes, i.e. square, circular, rectangular, elliptical, pentagon and hexagon are also one on the mechanical element which is being utilized in MEMS sensor/device arena in broad sense according to the application and/or specifications for fabrication of pressures sensors, accelerometers, gyroscopes, capacitive micromachined ultrasonic transducers (CMUT) and piezoelectric micromachined ultrasonic transducers (PMUT) [2], [28], [32], [61]. The cantilevers are very sensitive to pressure which are compatible with a small range of pressure measuring applications. In large pressure range measurement, then diaphragms of different shapes are a better option that plays a significant role in terms of the sensor's performance, sensitivity, and non-linearity [4], [8]. Three different diaphragm shapes, i.e. circular, square, and rectangular, for surface acoustic wave measurement in which the circular shape diaphragm-based sensors show maximum sensitivity [61]. The mathematical modelling, FEM-analysis, and comparison of elliptical capacitive pressure microsensor with circular capacitive pressure microsensor are presented [4], [8], [62]. The elliptical shape of the diaphragm is also utilized for the fabrication of the SiGe CMOS capacitive pressure sensor with the signal processing circuitry [63].

In paper electronics, different diaphragm shapes have also been utilized in designing pressure sensors for healthcare, acoustic pressure, and air pressure monitoring [28], [29], [32]. The square shape pressure sensor is presented as paper-like materials, i.e. Al-foil (used as parallel plate electrode of the sensor), double-sided tape (for clamping the edges of square pieces), a microfiber wipe, and sponge (for dielectric material) is utilized in which the sensor with air as the dielectric medium has maximum sensitivity among all these designs [29]. The circular, square, and rectangular shape of capacitive pressure sensors are fabricated using DIY and garage fabrication techniques made of Al coted Kapton foil and six-layer double-sided tapes for wheezing and asthma monitoring [32].

This chapter presents the design and analysis of five different shapes, i.e. circular, elliptical, square, rectangular, and pentagon of normal mode capacitive pressure sensor for air and acoustic pressure monitoring. First, paper-like materials, i.e. Al coted Kapton sheet and double-sided tape, are used for fabricating capacitive pressure sensors. Then sensors are characterized for acoustic and air pressure monitoring.

## 3.2 Governing equations

The maximum deflection in all different shapes of diaphragms that are clamped at the edges is given in **Table 1**.

The deflection equation is obtained from a partial differential equation which is being solved after considering multiple boundary conditions like deflection at the edges is zero, the slope of deflection at the edges is zero, deflection is maximum in the center of the diaphragm, and slope of the deflection at the center of diaphragm is zero. zero.

## 3.3 Fabrication

The fabrication of normal mode capacitive pressure sensor starts with cutting the Al-coated Kapton sheet (Liren's LR-PI 100AM of 25 µm polyimide coated with 200 nm aluminum) with the CO2 Laser tool, and the wave properties are kept the same, which is utilized to cut the rectangular/square shapes for designing cantilever based capacitive pressures sensors. The fabrication steps are shown in Figure 5-(a-e), which explains cutting, placing, and pasting, i.e., various techniques of the DIY technique.

**Table 1** Derived mathematical formula of maximum deflection in different shapes of the diaphragm.

| S. N. | Diaphragm shape | Maximum deflection | Corresponding parameters |
|---|---|---|---|
| **1.** | Circular | $\frac{Pr^4}{64D}$ | $r$ = radius |
| **2.** | Elliptical | $\frac{P}{8D}\left[\frac{a^4b^4}{3(a^2+b^2)+2a^2b^2}\right]$ | $a$ = semi-major axis<br>$b$ = semi-minor axis |
| **3.** | Pentagon | $0.0041\frac{Pa^4}{D}$ | $a$ = edge length |
| **4.** | Square | $0.00133\frac{Pa^4}{D}$ | $a$ = side length |
| **5.** | Rectangular | $0.00133\frac{Pa^4b^4}{D[7(a^4+b^4)+4a^2b^2]}$ | $a$ = side length<br>$b$ = side width |

The single-touch mode capacitive pressure sensor is fabricated using our approach requires

DIY-technique or garage fabrication process and low-cost materials like Al coted Kapton foil (Liren's LR-PI 100AM of 25 μm polyimide coated with 200 nm aluminum), double-sided scotch tape, and glass pieces (7.5 cm × 5 cm × 0.5 cm). The steps to fabricate sensors is as follows:

- **<u>Step-1</u>:**The Al coted Kapton sheet adheres on top glass substrate for cutting two pieces of specific shape from aluminum (Al) sputtered polyimide (PI) sheets from 2000 Carbon Dioxide ($CO_2$)laser tool**[Figure 3.1-(a)]**.

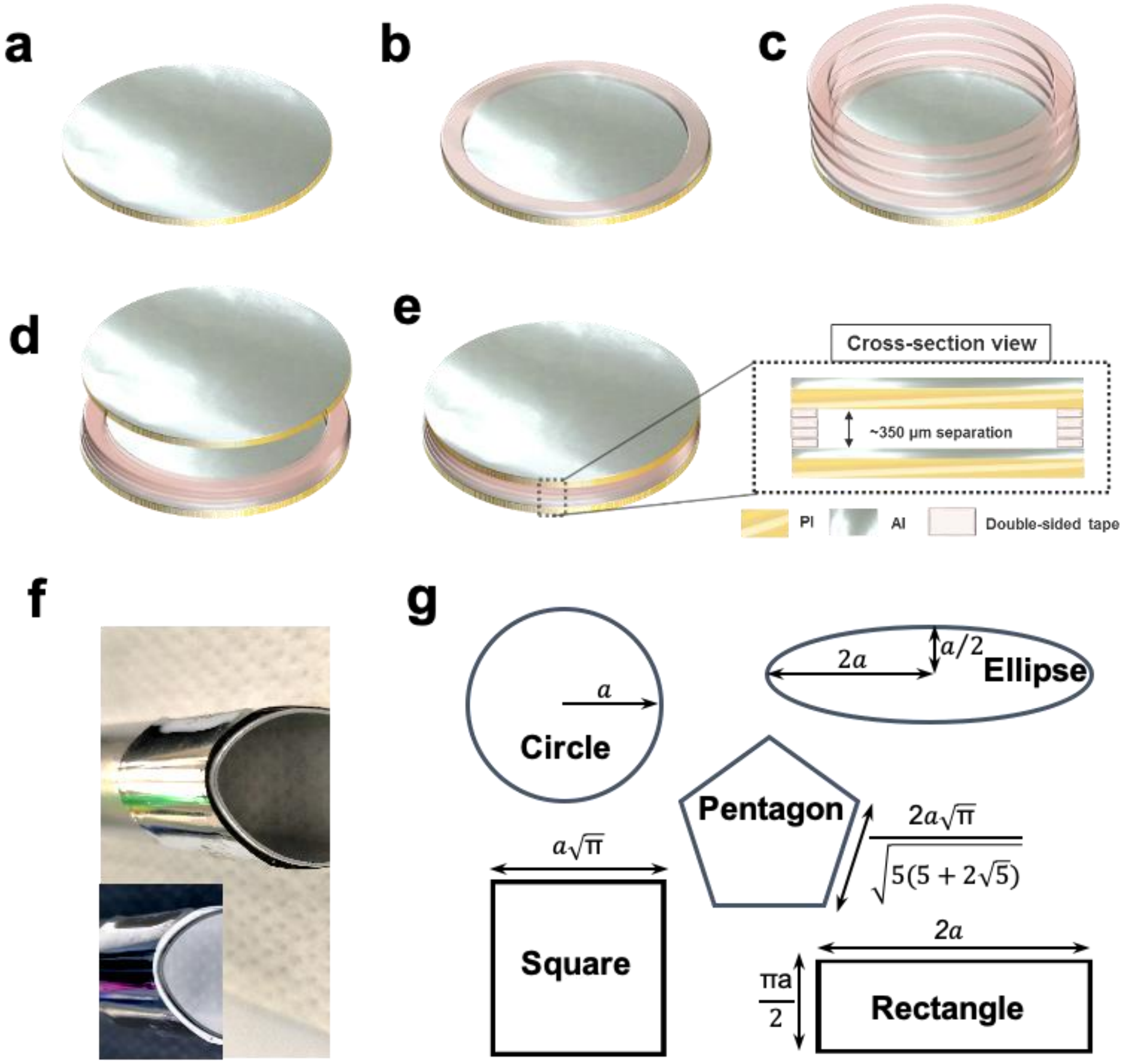

**Figure 5(a-e)**Schematic illustration of steps involved to fabricate diaphragm based normal mode capacitive pressure sensor for different five shapes in which **(f)** represents the final architecture of sensor to show the flexibility of sensor and **(g)** represents the shape and surface area of circle, ellipse, square, rectangle and pentagon. The notable point here is that the surface areas of all the diaphragms are kept constant.

- **<u>Step-2</u>:** One sample of Al coted Kapton sheet is kept adhered to on top glass, which will act asa carrier substrate, to fix the bottom electrode and provides mechanical strength to the capacitive structure. After removing the accessive film from another sample, the diaphragm is obtained which acts as the mechanical sensitive element for capacitive structure. The double-sided tape is used to clamp the edge of the mechanical sensitive element and then four more layersare placed on top of that **[Figure 5-(b-d)]**. The power, speed, vertical separation between laser and substrate, and the frequency is optimized for scribing which is 10%, 20%, 2 mm, and 1000 ppm, respectively.

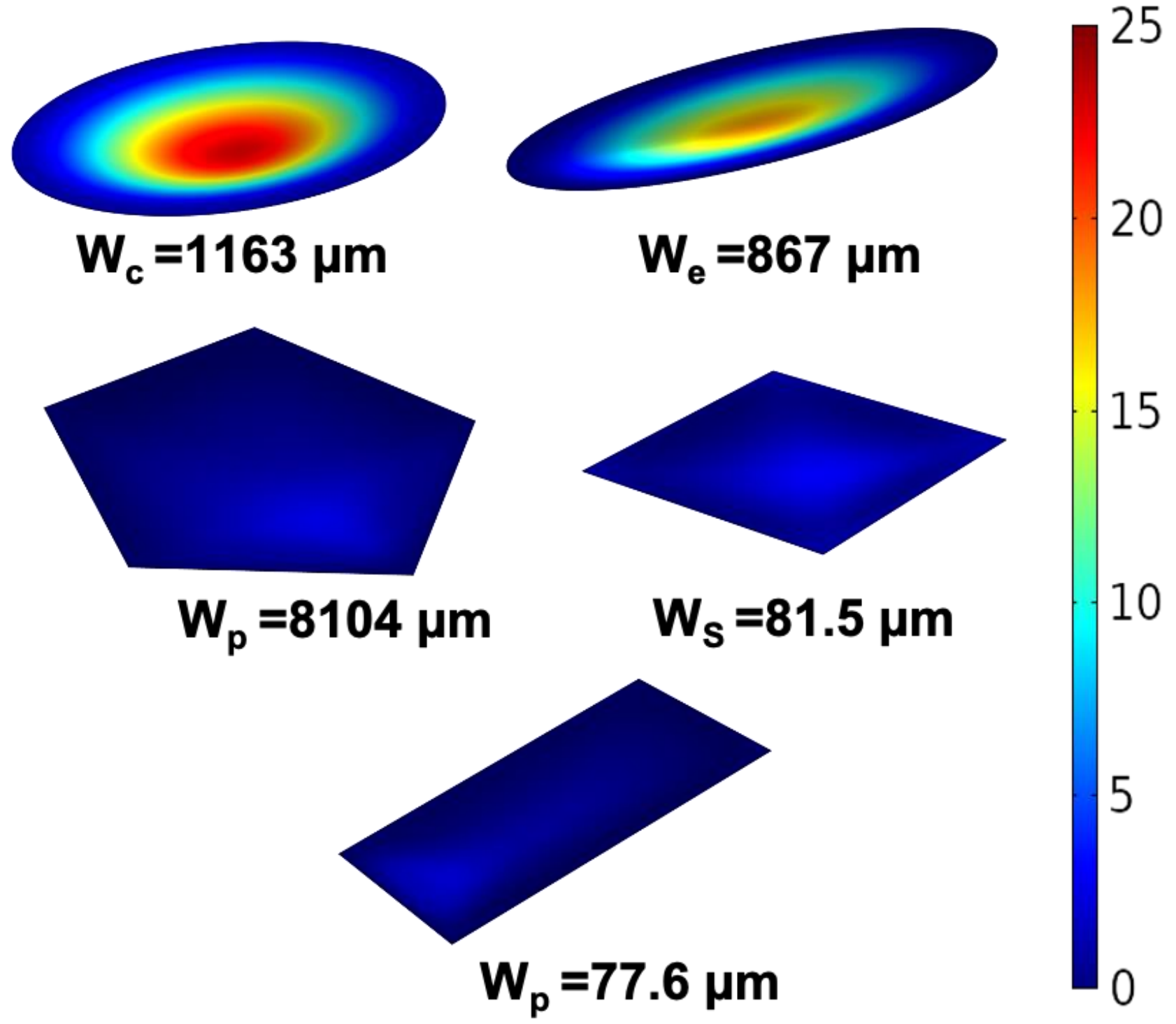

**Figure 6.** Finite element simulation of diaphragm deflection in five different shapes using COMSOL® Multiphysics.

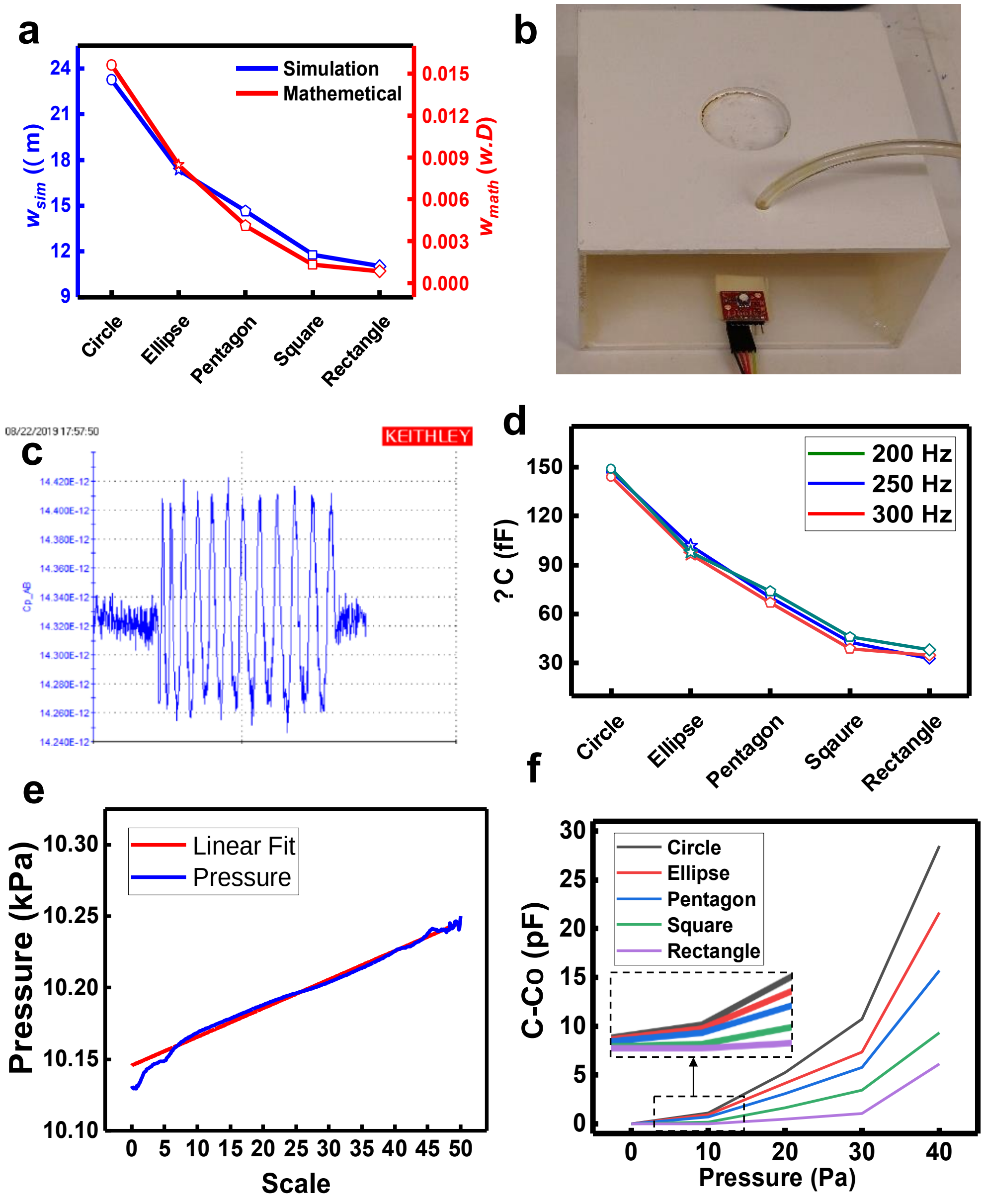

**Figure 7.** Experimental results of normal mode capacitive pressure sensor.**(a)** Comparison plot of deflection in the various shape of diaphragms which is obtained from mathematical modeling and finite element simulation. **(b)** Acrylic box to show the air-pressure monitoring set-up. **(c)** Output response for the circular diaphragm at 300 Hz. **(d)** Response of all different shapes of normal mode capacitive pressure sensors for acoustic pressure at 200 Hz, 250 Hz, and 300 Hz. **(e)** Linear fit of the pressure. **(f)** Response of all different shapes of normal mode capacitive pressure sensors for 0 – 40 Pa air pressure.

- **<u>Step-3</u>:**Place another piece (depends on the shape of the sensor is fabricated) placed on the layer of double-sided tape, which acts as the diaphragm. The air is a dielectric medium between parallel plates capacitors [Figure 5-e].

Figure 5-f shows the flexibility of the sensor is shown and Figure 5-g shows the shape and size of the diaphragm, which is mechanically sensitive and utilized for pressure detection in normal mode flexible capacitive pressures sensors.

## 3.4 FEM – Simulations

The COMSOL Multiphysics simulation tool is utilized to compare the mathematical analysis with finite element simulation. The finite element simulated values of deflection in various diaphragms are shown in **Figure 6**. The trend in diaphragm deflection, which is being obtained from mathematical modelling, is the same as the trend obtained from simulation, and the plot is shown in **Figure 7-a**.

## 3.5 Characterization

### 3.5.1 Experimental set-up and result due to acoustic pressure

To mimic the acoustic sensor setup conditions, we put all the diaphragm shapes under an equivalent sound pressure level (SPL) intensity of 94 dB, which is equivalent to a loud sound produced by human beings. To find the equal value of pressure is Pascal (Pa), we convert the value of Sound Pressure Level (SPL) into Pascal (Pa).

$$SPL(dB) = 20\, log_{10} \left| \frac{Measured\ Sound\ Pressure}{Reference\ Pressure} \right| \tag{3.1}$$

where the reference pressure was set to 20 μPa, which is the threshold of human hearing.45 An SPL of 94 dB corresponded to 1 Pa. The FEM simulation results to visualize the deflection behaviour in each diaphragm are shown in Figure 6. The deflection at the centre of each diaphragm against each shape is plotted in Figure 7-a. The mathematical values of deflection in terms of "D" ($w_{math} \times D$)for each shape are plotted in the same graph to verify that the simulation results follow the same performance trend as indicated by the respective equations. The identical diaphragms are then subjected to an equal pressure as used in the air pressure sensing experimental setup to observe the deflection response for enormous pressures (~40 Pa). The results of FEM simulations, along with maximum deflection, are shown in Figure 6. The trend remains the same for this more considerable pressure, albeit with a much larger deflection.

For the acoustic sensing experiment, we set up the experiment to evaluate the performance of each diaphragm for sound detection. We attached each of the sensors on a glass slide with the top metal plate (the diaphragm) of the sensor facing upwards, and the bottom plate is fixed to the glass slide. A Bluetooth speaker (JBL Go Portable Speaker) was placed at a 2 mm distance from the top metal diaphragm. Three different frequency sound tones (200 Hz, 250 Hz, and 300 Hz) were played through the speaker, and the capacitance was monitored using Keithley Semiconductor Characterization System (Model – 4200 SCS). The sound amplitude was set to 94 dB, which corresponds to a pressure of 1 Pa that was used in the FEM simulations. We get a sinusoidal output capacitance response and a sample output response for the circular diaphragm at 300 Hz shown in **Figure 7-c**. The results from three different input sounds are shown in **Figure 7-d**. Three different sound frequencies were chosen to make sure that the frequency of the input sound does not affect the result in any other way. Sound is inherently a sinusoidal wave by nature. The change in capacitance ($\Delta C$) was calculated for each diaphragm by calculating the peak-to-peak amplitude of the output observed for each sound frequency. It can be seen that the experimental results verify mathematical and simulation results. The circular diaphragm shows the largest deflection with 150 fF change in capacitance, followed by the elliptical diaphragm. The results for the elliptical diaphragm and the rectangular diaphragm are close to each other because the dimensions of the elliptical diaphragm (a = 2 cm, b = 0.5 cm) form a shape similar to that of the rectangular diaphragm.

### 3.5.2 Experimental set-up and results due to air pressure

We self-designed a set-up to exert pressure on the top electrode for the air pressure experimental set-up, a mechanically sensitive diaphragm. A 5 mm diameter plastic pipe was connected to the air valve. A custom scale was made on the valve with 0 – 50 such that the value is 0 for a fully closed valve and 50 for a fully opened valve. The end of the pipe was then inserted into a hole inside the top layer of an acrylic box such that the air coming out of the nozzle will apply pressure on the bottom surface of the acrylic box **[Figure 7-b]**. A commercial pressure sensor (MS5803-14BA) was then attached to the bottom surface right below the pipe's opening. This MEMS pressure sensor measures the absolute pressure of the fluid around it, which includes air, water, and anything else that acts like a viscous fluid. The valve was opened from 0 to 50 with increments of 5, and the corresponding pressure was measured using the air pressure sensor. A linear fit of the pressure observed corresponding to the value of scale can be seen in Figure 7-e. This plot is used to convert the value of the custom scale to a pressure value in Pascal (Pa). To experiment, we placed all of the diaphragms at the same place as the pressure sensor. We opened the valve from 0 to 24 (after 24, the increased pressure makes the deflection difference too large for a graphical representation). The change in capacitance from the starting value for each diaphragm is plotted against the applied pressure, and the results are plotted in Figure 7-f. The experiment further verifies that the diaphragm exhibits a similar deflection response as expected from the mathematical analysis and the FEM simulations. The circle diaphragm undergoes the most considerable deflection, followed by the ellipse diaphragm. The rectangular shape capacitive pressure diaphragm shows the minor deflection. In terms of linear response Figure, 7-f confirms that circular shape capacitive pressure sensor shows maximum non-linearity in output response; however, the rectangular diaphragm provides the most linear response. Therefore, if we are looking for a susceptible sensor, we should go with circular shape diaphragms; however, a rectangular shape sensor should be preferred for linear response.

## 3.6 Conclusion

In this chapter, the normal mode of capacitive pressure sensors with different diaphragm shapes (i.e. circular, elliptical, square, pentagon, and rectangular) are presented. The mathematical and FEM simulation results are verified with two experimental setups, i.e. acoustic pressure and air pressure. After characterization, it is found that the circular shape diaphragm deflects maximum among the five shapes, which mean circular shape capacitive pressure sensor shows maximum mechanical sensitivity. If sensitivity is essential for a required application, a circular or less elliptical diaphragm shape should be chosen. However, the circular shape capacitive sensor response is highly non-linear than the other five shapes of diaphragms. We have used low-cost, recyclable materials, which can result in reduced financial and environmental costs. The elliptical-shaped diaphragm has less material wastage while having a comparable performance in comparison with circular-shaped diaphragms. For scalable manufacturing techniques with the least material wastage, a square-shaped diaphragm is more beneficial. In addition, square-shaped diaphragms show a linear response. The response becomes non-linear as we move toward circular-shaped diaphragms.

# Chapter – 4

# Singleanddouble touch mode capacitive pressure sensors

## 4.1 Introduction

Apart from different cantilever/diaphragm shapes based on capacitive pressure sensors, the modes also play an essential role in pressure measurement. For example, the cantilever or diaphragm-based normal mode capacitive pressure sensors are very suitable for a small range of pressure measurement; however, if we try to measure the large range of pressure, the sensors will not be the highly sensitive and non-linear response.

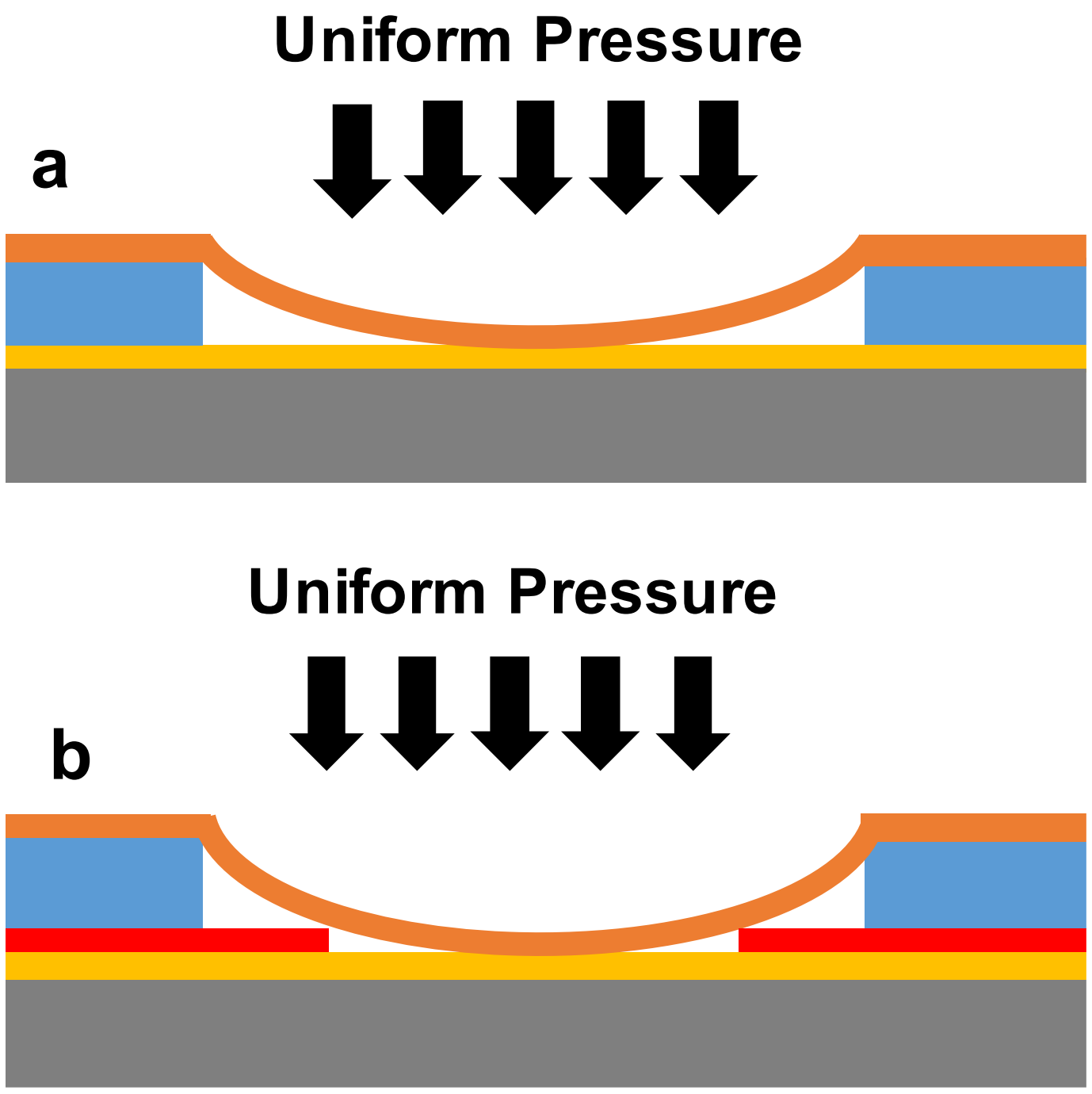

**Figure 8.**Schematic to (a) single and (b) double touch mode capacitive pressure sensor.

To enhance the properties of the capacitive pressure sensor, further analysis is proposed according to the modes of the capacitive pressure sensor in MEMS and circular shape based single and double touch mode capacitive pressure sensors. In a single touch mode capacitive pressure sensor **[Figure 8-a]**, the top electrode, i.e., mechanically sensitive diaphragm, touches the bottom electrode, separated by a thin dielectric layer and the pressure at which the diaphragm touches, called touch point pressure. The flexible touch mode capacitive pressure sensor is proposed by Prof. A. Vijayaraghavan group at the National Graphene Institute, University of Manchester. In this work, the graphene-polymer heterostructure is suspended. It acts like a pressure-sensitive diaphragm. The authors reported all four different regions (i.e., normal, transition, linear, and saturation) for the pressure range of 0.5 - 8.5 kPa with the sensitivity of 14.8 $\mu$V/Pa, in the linear region.

Moreover, in the double-touch mode capacitive pressure sensor **[Figure 8-b]**, the mechanically sensitive diaphragm is separated in such a way so that we get two touchpoints. Another layer will have a small hole; first, the diaphragm touches the bottom electrode, and that pressure is known as first touchpoint pressure. However, when the same diaphragm touches another thin dielectric with a small hole that provides second touchpoint pressure. The advantages of touch mode capacitive pressure sensor (single and double both) have multiple following properties over the normal mode of capacitive pressure sensors [30], [31], [64], [65]:

- Robust structure
- Suitable for extensive pressure range monitoring
- Provides linear response in the specific pressure range

A single touch mode capacitive pressure sensor provides the properties mentioned above;

however, it saturates very fast. To overcome this disadvantage, H. Lv et al. presented a MEMS-based double-touch mode capacitive pressure sensor, which increases the linear response of output range for large pressure range measurement [66].

Herein, the single touch mode capacitive pressure sensor is fabricated and experimentally characterized first. Then, to overcome the challenges mentioned above, the double touch mode capacitive pressure sensor is also fabricated and experimentally characterized. Finally, household materials like Al coated polymer sheet, double-sided, and Kapton tape is utilized for fabrication using garage fabrication or DIY - technique.

## 4.2 Governing equations

The governing partial differential equation of deflection in diaphragm is given by partial differential equation:

$$\left(\frac{\partial^2 W}{\partial r^2}\right)^2 + \frac{1}{r^2}\left(\frac{\partial W}{\partial r}\right)^2 + \frac{1}{r}\frac{\partial^2 W}{\partial r^2}\left[2\frac{\partial W}{\partial r} - \frac{h}{D}\frac{\partial \varphi}{\partial r}\right] = \frac{P}{D} \quad (4.1)$$

where, h, D, Φ, and W are diaphragm thickness, the flexural rigidity of diaphragm, airy stress, and diaphragm deflection at radius r respectively. The flexural rigidity is a function of young's modulus of elasticity, diaphragm thickness, and Poisson's ratio of diaphragm material. If the circular diaphragm, made of elastic, homogeneous, and isotropic material, is clamped at the edges, then after applying the boundary conditions, the deflection in the diaphragm at any distance r due to uniform pressure application is given by:

$$W(r) = W_0\left[1 - \left(\frac{r}{R}\right)^2\right]^2 \quad (4.2)$$

where $W_0$ is the diaphragm deflection at the center and R is the diaphragm radius. A large deflection due to the application of pressure in the circular diaphragm is given by:

$$W_{0,l} = \frac{PR^4}{64D}\left[\frac{1}{1 + 0.488\ \frac{w_{0,l}^2}{h} + \frac{\sigma h R^2}{16D}}\right] \quad (4.3)$$

where, h, σ, D, P are diaphragm thickness, build-in-stress, flexural rigidity, and applied pressure respectively. A small deflection in circular diaphragm with build-in stress is given by:

The base capacitance of the parallel plate capacitive pressure sensor is given by:

$$C_0 = \frac{\varepsilon_0\ \varepsilon_r A}{d} \quad (4.4)$$

where $\varepsilon_0$ is the permittivity of air, $\varepsilon_r$ is the permittivity of the medium. The capacitance variation due to the application of pressure on the circular diaphragm (for both small and large deflection) is given by:

$$C_w = \int_0^{2\pi}\int_0^{a} \frac{\varepsilon_0 \varepsilon_r r dr d\theta}{d\ -\ W(r)} \quad (4.5)$$

## 4.3 Fabrication

The section is divided into two subsections which explain the fabrication of single and double touch mode capacitive pressure sensors.

### 4.3.1 Single touch mode capacitive pressure sensor

The The single-touch mode capacitive pressure sensor is fabricated using our approach requires DIY-technique or garage fabrication process and low-cost materials like Al coted Kapton foil (Liren's LR-PI 100AM of 25 ¬µm polyimide coated with 200 nm aluminium), double-sided scotch tape, and glass pieces (7.5 cm √ó 5 cm √ó 0.5 cm). The steps to fabricate sensors is as follows:

- Step-1: The Al coted Kapton sheet has adhered to the top glass substrate for cutting two pieces of circular shape from aluminium (Al) sputtered polyimide (PI) sheets (radius=1.2 cm) from 2000 Carbon Dioxide (CO2) laser tool (Universal Laser Systems PLS6.75) **[Figure 9-a]**.

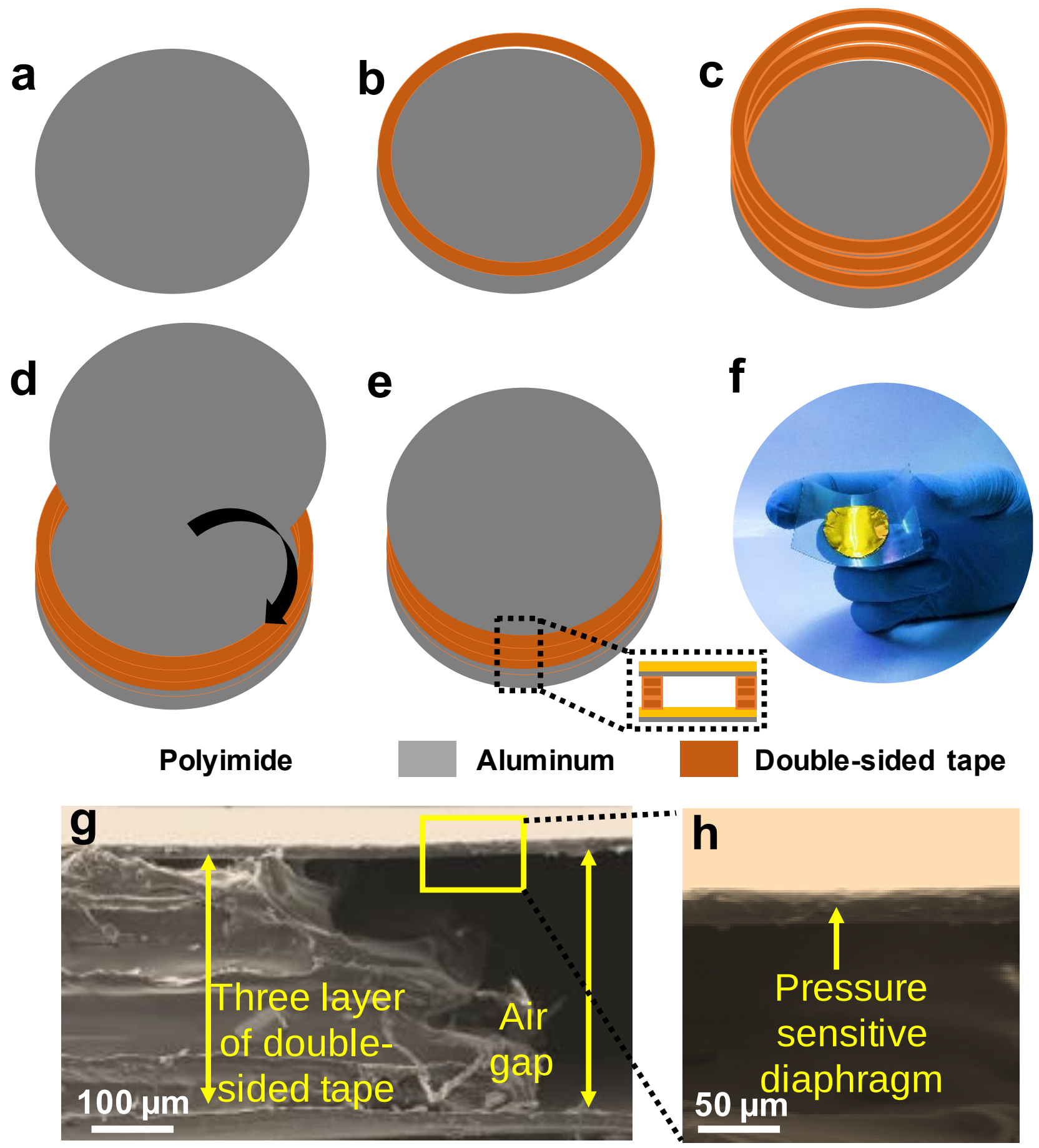

**Figure 9**Schematics of the process flow for fabrication of polymer/paper-based touch mode pressure sensors. **(a)** Cut the two pieces circular pieces from 25 µm thick Al coated PI sheet for the top and bottom electrode. **(b)** Place a single layer of double-sided tape ring to clamp the edges. **(c)** Repeat

the number of layers depending upon the needed airgap (here we have three layers of double-sided tape). **(d)** Adhere the top electrode to the third layer with double-sided tape. (e) Schematic of the fabricated capacitive pressure sensor. **(f)** Photographic image of fabricated capacitive pressure sensor on a Polyethylene Terephthalate (PET) sheet to shown to admit the flexibility of sensor under the compressive stress. **(g)** SEM image of pressure sensor showing three layers of tape and air gap. **(h)** The zoomed part shows the diaphragm of the sensor which is a pressure-sensitive mechanical element.

- **Step-2:**One sample of Al coted Kapton sheet is kept on top glass, which will act as carrier substrate to fix the bottom electrode and provide mechanical strength to the capacitive structure. After removing the accessive film from another sample, the diaphragm is obtained, which acts as the mechanically sensitive element for capacitive structure. The double-sided tape is used to clamp the edge of the mechanically sensitive element, and then two more layers are placed on top of that **[Figure 9-(b-c)**]. The power, speed, vertical separation between laser and substrate, and frequency are optimized for scribing, 10%, 20%, 2 mm, and 1000 ppm, respectively.
- **Step-3:**Place another circular piece on the double-sided tape layer, which acts as the diaphragm, and the air is the dielectric medium between parallel plate capacitors [Figure 9-(d-e]. The photographic image of the actual sensor is shown in [Figure 9-f]. The SEM images of the cross-sectional area have been taken **[Figure 9-(g-h)]** to achieve the value of the separation gap, which is found to be almost 400 $\mu$m.

### 4.3.2 Double touch mode capacitive pressure sensor

The double touch mode capacitive pressure sensor is fabricated in the same way as above; however, one more dielectric layer, Kapton-tape, has one small concentric circular hole of 5 mm.

The fabrication steps are shown in **Figure 10-(a-f)**, which are almost the same as the fabrication steps of single-touch mode capacitive pressure sensor except for the dielectric layer (Kapton-tape) with a small hole of 5 mm diameter, which is being confirmed from the SEM-images which are taken and shown in **Figure 10-(g-h)**.

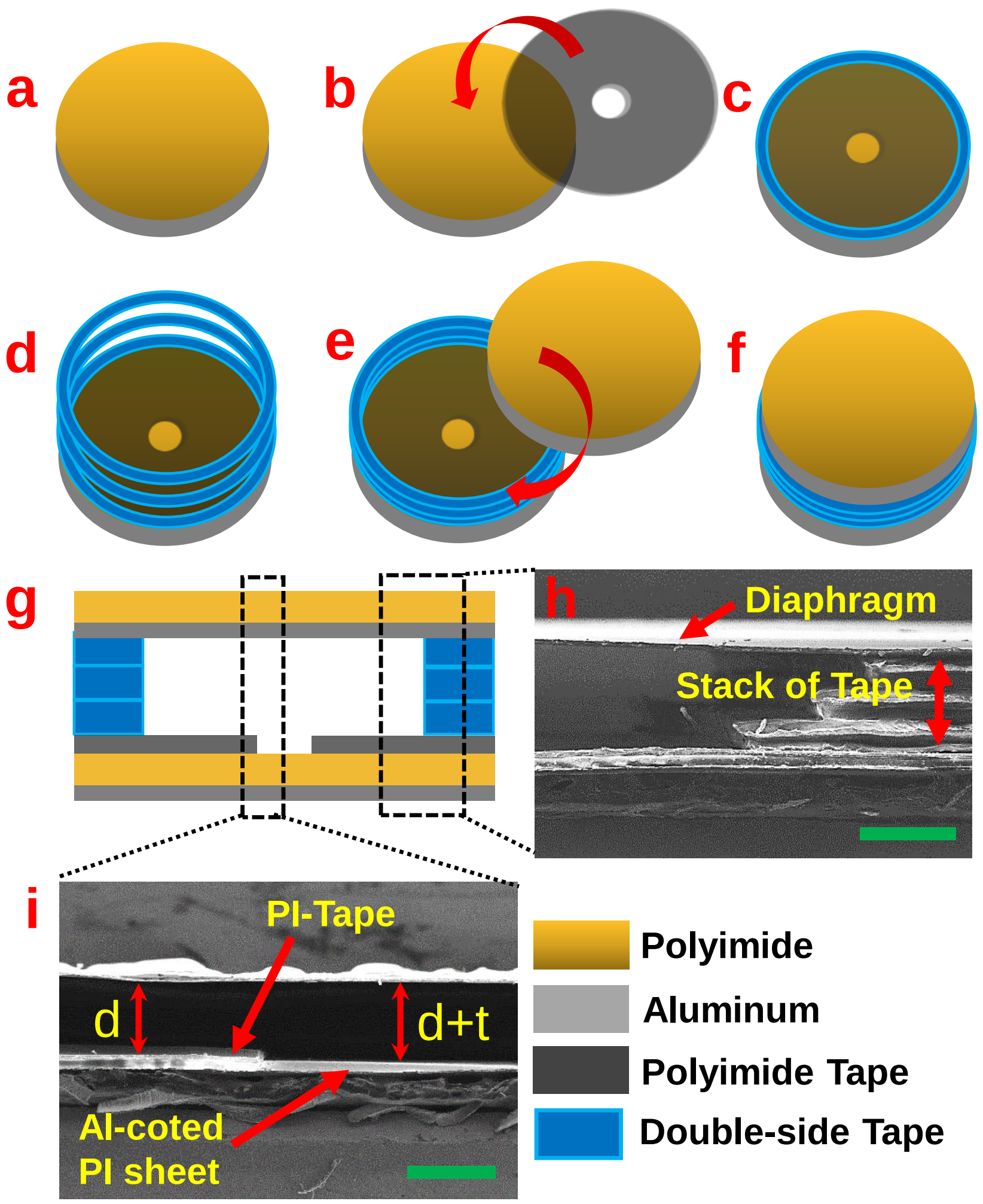

**Figure 10.** Figure 4.2.2. Schematic diagram of the fabrication process of sensors. (a) Laser cutting the circular shape from the Al coted PI sheet of 2.6 cm diameter. (b) Placing the PI tape with 2.6 cm of external diameter and the 5 mm diameter of inner circular cut is placed on the PI side of the previous circular shape. (c) Placing the single and layer of double-sided at the edges. (d) Repeating

the previous steps two more times. (e) Placing the circular shape, which can obtain from the first step, on top of the third layer of double-sided tape. (f) Final illustrated image of sensor. (g) Cross-sectional view of the sensor. (h) SEM image of the cross-sectional view of the sensor. (i) SEM image of the half middle part to show the second cavity at which the mechanical sensitive diaphragm touches first after pressure application. The bar scale of both the SEM images is 500 μm. The thickness of the PI tape is found to be 15 μm, d = 425 μm, d+t = 451 μm because of the adhesiveness of the PI tape. The thickness of the single double-sided tape is found to be 110 μm. The diaphragm is bent at the center due to pre-stress/build-in stress and gravity effect.

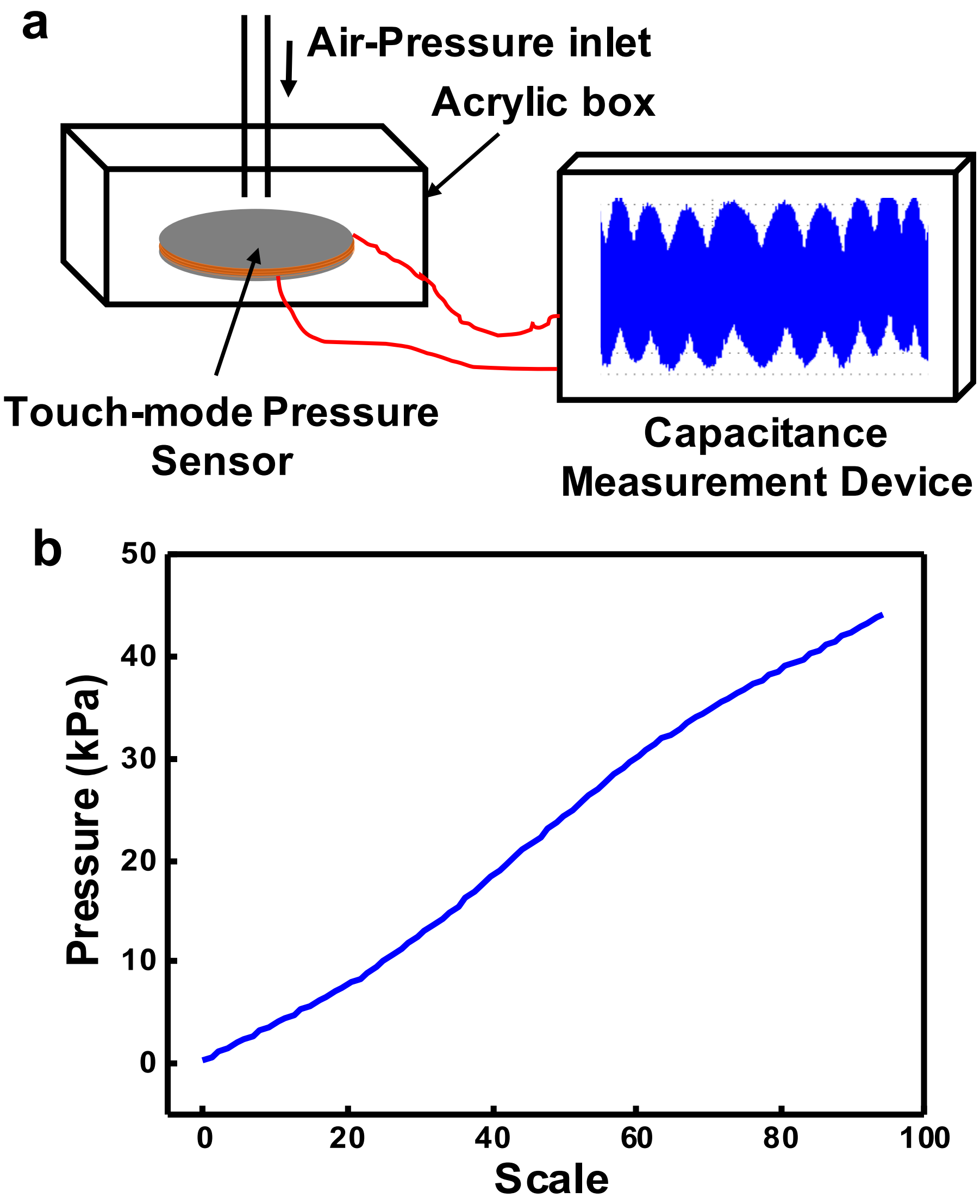

**Figure 11**Schematic to **(a)** air pressure set-up. **(b)** Response of commercially available MEMS capacitive pressure sensor (MS5803 – 14BA).

## 4.4 Characterization

### 4.4.1 Experimental set-up

Anexperimental setup is manually designed to reach high pressures on the diaphragm to achieve the single and double touch mode and utilized to experimentally characterize the fabricated capacitive pressure sensors. We placed both sensors one by one on the bottom of an acrylic box (available in the laboratory) with a small hole of about 1 cm diameter. An air pressure nozzle was inserted inside that hole on the top of the acrylic box [Figure 11-a]. The outlet of the nozzle is directly above the mechanically sensitive diaphragm of the fabricated capacitive pressure sensor. A custom scale from 0 – 100 was drawn on the knob of the nozzle to find out how much pressure the nozzle can apply to each value on the scale. A commercial MEMS capacitive pressure sensor (MS5803 – 14BA) was used to calibrate the value of the custom scale to the corresponding value of pressure. The commercial pressure sensor was placed under the opening of the nozzle where the sensor has to be placed. The knob was opened from 0 – 100 while noting down the pressure values from the commercial sensor. The response of the commercially available MEMS pressure sensor is shown in Figure 11-b. The knob of the air source results in a linear increase in pressure.

### 4.4.2 Experimental result of single touch mode capacitive pressure sensor

The fabricated sensor was placed under the nozzle, and the knob was opened from 0 – 100 while taking the reading at intervals of 10 **[Figure 11-a]**. It can be experimentally observed that the fabricated single touch mode pressure sensor **[Figure 12]** provides a non-linear response in the pressure range of 1 – 8 kPa. This is expected from a capacitive sensor in the normal range of operation as long as the deflection in the mechanically sensitive diaphragm of the pressure sensor is less than 1/3rd of the separation gap and follows Kirchhoff's plate theory of deflection. As we reach higher pressure values (more than 8 kPa), the sensor goes into transition mode (8–10 kPa) of operation. In this mode, the pull-in phenomena occur, and the mechanically sensitive diaphragm touches the backplate of the sensor, which is fixed. After that, a further increase in pressure (more than 10 kPa), transfers the sensor into touch mode operation. After that, the response becomes linear for a wide range (from 10 – 40 kPa), and the capacitance increases as the large area of the diaphragm touch the backplate. After that, the response saturates as the maximum portion of the diaphragm is now stuck to the bottom plate. plate.

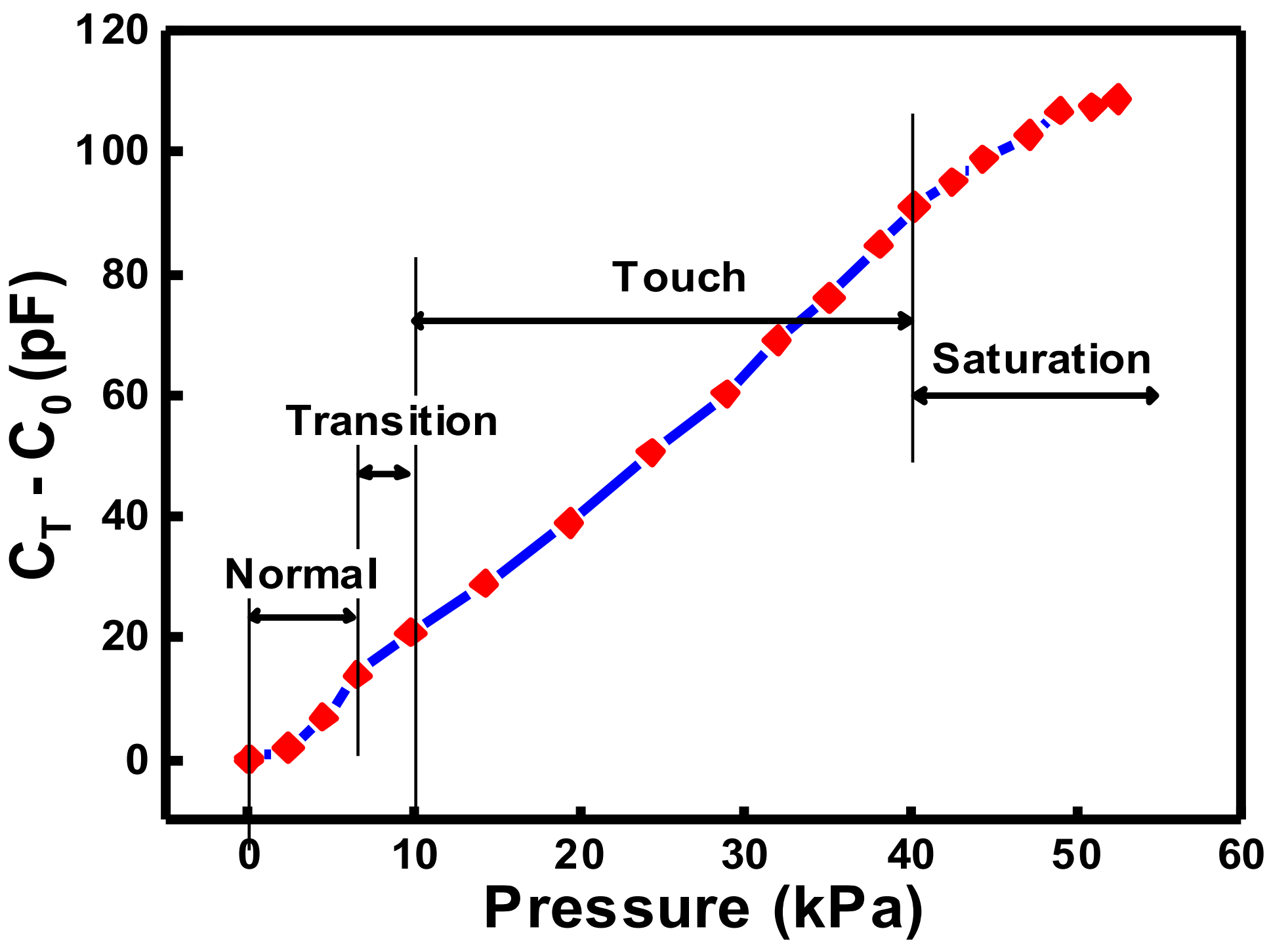

**Figure 12.**Response of fabricated single touch mode capacitive pressure sensor.

### 4.4.3 Experimental result of double touch mode capacitive pressure sensor

The response of the sensor started with nonlinear behavior when the sensor was operating in the normal region (shown in red) for the 0 – 7.5 kPa range [Figure 13]. At a pressure of 7.5 kPa, the pull-in phenomena occurred, and the diaphragm touched the bottom electrode. This pressure is known as the first touch pressure point (TP1). At $TP_1$, the sensor entered the transition region (for pressure more than 7.5 kPa), in which the diaphragm touched the bottom cavity, which is at a depth of. As the pressure increased to 9.7 kPa, the diaphragm touched the PI tape with a small concentric hole. This pressure is known as the second touch pressure point ($TP_2$). The fabricated capacitive pressure sensor started operating in the linear region from 14.24 kPa, which is our range of interest for designing the integrated circuitry for any efficient and high precision application. Beyond the 54.9 kPa pressure point, the sensor operated in the saturation region because the maximum area of the mechanically sensitive diaphragm was already touching the bottom. In the linear region, the sensor has a sensitivity of 0.674 fF/Pa.

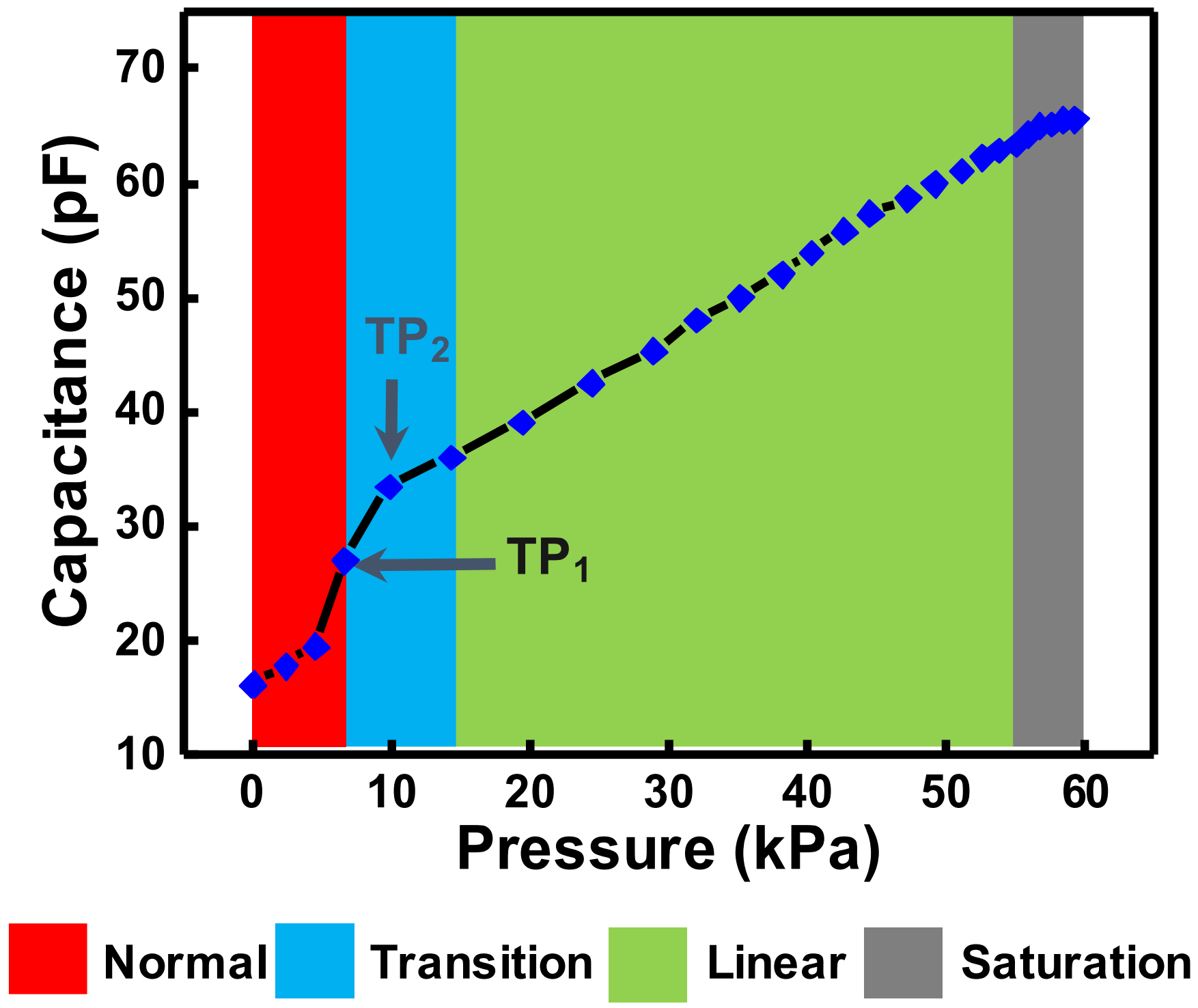

**Figure 13.**Response of fabricated double touch mode capacitive pressure sensor.

## 4.5 Conclusion

In this chapter, the single and double touch mode of capacitive pressure sensors are fabricated from household materials, i.e., Al coated Kapton sheet and double-sided tape, and experimentally characterized using air-pressure set-up. In double touch mode, the pressure sensitivity is almost the same as the touch mode capacitive pressure sensor. However, the linear operating range for the specified pressure range is increased for the same diaphragm radius. As a result, the linear range of operating pressure range for touch mode is 10 – 40 kPa which is 14.24 – 54.9 kPa for double touch mode capacitive pressure sensor.

# Chapter – 5

# Conclusion and Future Work

In the presented thesis, I explored cantilever, normal mode (with different diaphragm shapes), single and double touch mode capacitive pressure sensors. In chapter two, i.e., cantilever capacitive pressure sensor, I explored the design analysis, mathematical modelling, finite element simulations (using COMSOL Multiphysics and CoventorWare®) with fabrication and characterization and application for extensive range of pressure variation. After exploring different shapes and design of sensors, this research work concludes that as we increase the length of the cantilever, the sensitivity increases; however, response time decreases and resonates at the low frequency. However, the cantilever sensor designs do not apply to non-curvilinear surfaces and susceptible to noise. Therefore, the diaphragm shape designs based on normal mode capacitive pressure sensors are characterized and discussed the sensitivity and linearity of sensors; however, less applicable for high-pressure monitoring. Moreover, the circular shape capacitive pressure sensor shows maximum sensitivity; however, the response will be highly non-linear than other shape-based normal mode capacitive pressure sensors. After that, a single touch mode capacitive pressure sensor is fabricated and experimentally characterized. The touch mode capacitive pressure sensor is linear however saturates very fast. Therefore, a double touch mode pressure sensor is fabricated and experimentally characterized to increase the linear regime.

Herein, the capacitive pressure sensor is utilized to design the flexible capacitive pressure sensor using household materials using DIY-techniques for fabrication. The advantage of capacitive pressures is low-temperature drift, high sensitivity, and ease in fabrication; however, it requires a large die-area. Therefore, other techniques (like piezo-resistive sensing) can be utilized and explored a bit more for the design criteria and shape analysis and can be utilized for multiple applications like controlling drones, healthcare monitoring, consumer and portable electronics, and robotics [67 – 70]. The advantage of paper and paper-like materials like easy availability, low-cost and biodegradability might be a very fantastic option over the conventional electronic materials. Therefore, it might help to proceeds with a big step towards democratized [71-72] and freeform [73-74] electronics that enable smart living for all.

# Publications

Related JournalPublications:

- R. B. Mishra, N. El-Atab, A. M. Hussain, M. M. Hussain, “Recent Progress on Flexible Capacitive Pressures Sensors: From Design & Materials to Applications,” *Adv. Mater. Technol.*, vol. 6, no. 4, p. 2001023, 2021. [Link] [Cover article]
- R. B. Mishra, S. F. Shaikh, A. M. Hussain, M. M. Hussain, “Metal coated polymer and paper-based cantilever design and analysis for acoustic pressure sensing,” *AIP Adv.*,vol. 10, no. 5, p. 055112,2020.[Link]
- S. M. Khan*, R. B. Mishra*, N. Qaiser, A. M. Hussain,M. M. Hussain, “Diaphragm shape effect on the performance of foil-based capacitive pressure sensors,” *AIP Adv.*,vol. 10, no. 5, p. 015009, 2020. (*equal first author) [Link]

Related Conference Publications:

- R. B. Mishra, W. Babatain, N. El-Atab, A. M. Hussain, M. M. Hussain, “Polymer/paper-based double touch mode capacitive pressure sensing element for wireless control of robotic arm,”*15th IEEE International Conference on Nano/Micro Engineered Molecular Systems (IEEE-NEMS)*, pp. 95-99, San Diego, CA, USA, 2020. [Link]
- R. B. Mishra, S. M. Khan, S. F. Shaikh, A. M. Hussain, M. M. Hussain, “Low-cost foil/paperbased touch mode pressure sensing element as artificial skin module for prosthetic hand,” *3rd Inter. Conf. on Soft Robotics (RoboSoft)*, pp. 194-200, Yale University,CT, USA, 2020. [Link]

- R. B. Mishra, S. R. Nagireddy, S. Bhattacharjee, A. M. Hussain, "Theoretical Modelling and Numerical Simulation of Elliptical Capacitive Pressure Microsensor," *2nd IEEE Conference on Modeling of Systems Circuits and Devices*, pp. 17-22,Hyderabad, India, 2019. [Link]

Other Journal Publications:

- A. Fatema, S. Poondla, R. B. Mishra, A. M. Hussain, "A Low-Cost Pressure Sensor Matrix for Activity Monitoring in Stroke Patients using Artificial Intelligence" *IEEE Sensors J.*,vol. 21, no. 7, pp. 9546 – 9552, 2021.
- S. R. Nagireddy, K. S. C. Karnati, R. B. Mishra, A. M. Hussain, "Modelling of Multilayer Perforated Electrodes for Dielectric Elastomer Actuator Applications," *MRS Adv.*, vol. 14-15, pp. 765-771, 2020. [Link]

Other Conference Publications:

- S. Bhattacharjee, R. B. Mishra, D. Devendra, A. M. Hussain, "Simulation and Fabrication of Piezoelectrically Actuated Nozzle/Diffuser Micropump," *IEEE SENSORS*, pp. 1-4, Montreal, QC, Canada, 2019. [Link]
- K. S. C. Karnati, S. R. Nagireddy, R. B. Mishra, A. M. Hussain, "Design of Micro-heaters Inspired by Space Filling Fractal Curves,"*IEEE Region 10 Symposium (TENSYMP)*, pp. 231-236, Kolkata, India, 2019. [Link]
- S. R. Nagireddy, R. B. Mishra, K. S. C. Karnati, A. M. Hussain, "Modelling of Multilayer Perforated Electrodes for Dielectric Elastomer Actuator Applications," *2nd IEEE Conf. on Modeling of Systems Circuits and Device*, pp. 34-38, Hyderabad, India, 2019. [Link]